# Miniaturizing Color-Sensitive Photodetectors via Hybrid Nanoantennas towards Sub-micron Dimensions

Jinfa Ho[1,†], Zhaogang Dong[1,2,†,*], Hai Sheng Leong[1], Jun Zhang[1], Febiana Tjiptoharsono[1], Soroosh Daqiqeh Rezaei[3], Ken Choon Hwa Goh[1], Mengfei Wu[1], Shiqiang Li[1], Jingyee Chee[1], Calvin Pei Yu Wong[1], Arseniy I. Kuznetsov[1], and Joel K. W. Yang[1,3,*]

[1]Institute of Materials Research and Engineering, A*STAR (Agency for Science, Technology and Research), 2 Fusionopolis Way, #08-03 Innovis, 138634 Singapore

[2]Department of Materials Science and Engineering, National University of Singapore, 9 Engineering Drive 1, 117575, Singapore

[3]Singapore University of Technology and Design, 8 Somapah Road, 487372, Singapore

*Correspondence and requests for materials should be addressed to J.K.W.Y. (email: joel_yang@sutd.edu.sg) and Z.D. (email: dongz@imre.a-star.edu.sg).

**Abstract**

Digital camera sensors utilize color filters on photodiodes to achieve color selectivity. As color filters and photosensitive silicon layers are separate elements, these sensors suffer from optical cross-talk, which sets limits to the minimum pixel size. In this paper, we report hybrid silicon-aluminum nanostructures in the extreme limit of zero distance between color filters and sensors. This design could essentially achieve sub-micron pixel dimensions and minimize the optical cross-talk originated from tilt illuminations. The designed hybrid silicon-aluminum nanostructure has dual functionalities. Crucially, it supports a hybrid Mie-plasmon resonance of magnetic dipole to achieve the color-selective light absorption, generating electron-hole pairs. Simultaneously, the silicon-aluminum interface forms a Schottky barrier for charge separation and photodetection. This design could potentially replace the traditional dye-based filters for camera sensors at ultra-high pixel densities with advanced functionalities in sensing polarization and directionality, as well as UV selectivity *via* interband plasmons of silicon.

**Teaser: Hybrid plasmon-Mie resonance enables color-selective photodetectors, towards optical diffraction limit for next generation CMOS sensors.**

Miniaturization of CMOS color sensors based on silicon (Si) technology could enable high-resolution light-field sensor arrays,[1] hyperspectral imaging,[2] and miniaturized spectrometers.[3-5] However, due to its broadband absorption in the visible spectrum, Si-based photodetectors are intrinsically "color blind". Thus, color filters are required to integrate with the detectors to achieve spectral sensitivity. Typical color filters consist of conventional dye-based red-green-blue (RGB) color filters, linear variable filters,[6-8] Fabry-Perot etalon bandpass filters[9, 10] and so on. However, an inherent drawback in these approaches is the optical cross-talk that occurs when sensors are placed closely together. Cross-talk occurs when light that passes through one color filter is undesirably absorbed by surrounding photodetector elements. Although this problem can be mitigated *via* back-side illumination (BSI) detector technology by decreasing the distance between color filters and the light-absorbing Si layer beneath, parallax effects limit the degree of miniaturization of optical detectors to dimensions larger than a few micrometers.

Recently, there has been progress in exploring various designs to miniaturize spectrally selective photodetectors, e.g. utilizing nanowires,[3, 11-15] large array filters,[2, 4, 16] plasmonic hot-electrons,[17-23] graphene nanoribbons,[24, 25] colloidal quantum dot absorption,[26] photonic crystals,[27, 28] Schottky barrier for near-IR regime,[17, 18] and ITO-Si junction with scanning probe setup.[29] However, these approaches are ill-suited for scale up and/or incompatible with CMOS fabrication processes. On the other hand, other approaches based on tall Si nanowires, incorporating *p-i-n* junctions,[5, 30] can achieve the miniaturized spectrometer functionalities due to its waveguide mode resonance. Nevertheless, this approach is generally limited by its potential scalability and miniaturization. Alternatively, other CMOS compatible color sensor designs are based on *p-i-n* anti-Hermitian Si metasurface[31] or the Al-Si junction with a large pixel size of ~10 μm.[19] In addition, both of these designs[19, 31] exhibit significant dark current, which induces a large background signal for photodetection.

In this paper, we explore the extreme limit of decreasing the distance between color filter and sensor to zero, by nano-patterning light-absorbing silicon-aluminum (Si-Al) layer that imparts color sensitivity to Si. This design could essentially push pixel dimensions to the sub-micron regime and minimize cross-talks between adjacent color channels, originated from tilt illuminations. The hybrid Si-Al nanostructures support a hybrid plasmon-Mie resonance to achieve intrinsic color sensitivity. The ability to support magnetic resonances and their low optical losses make high-index semiconductors, such as Si, particularly attractive for applications in wavelength selective photodetection. Magnetic resonances[32] exhibit field confinement within the dielectric nanostructure and confers benefits for photocurrent generation by increasing the effective optical path of light in the semiconductor. Indeed, sharp spectral resonances in sub-wavelength Si nanodisks have been demonstrated due to low optical losses, indicating their suitability as miniaturized spectral filters. As an added advantage, Si devices and nanofabrication processes are well established and understood, allowing for solutions based on Si nanostructures to be easily scaled up and implemented in commercial products, as well as towards full UV-Vis multi-spectral photometers leveraging on interband plasmonic characteristics of Si nanostructures in the UV.[33]

**Results**

Figure 1a presents the schematic illustration of the designed color-sensitive detector, which consists of Si nanoposts with Al disks on top and Al film between the posts. Using finite-difference time-domain (FDTD) simulations, we obtained a suitable design consisting of nanodisk arrays with a height $h$ of 200 nm and pitch $\Lambda$ of 200 nm, used throughout the manuscript. The diameter

*D* of the Si nanodisk array is varied to achieve the desired color sensitivity for red, green, and blue. A 30-nm Al layer was then deposited using electron-beam evaporation. This Al layer has the following functionalities: The Al disk on Si nanostructure functions as a hybrid plasmon-Mie resonator, enabling enhanced absorption at the selected light wavelength. Simultaneously, the Al film forms a Schottky barrier with the *p*-doped Si substrate as the holes from the *p*-Si region diffuse towards Al to reach an equilibrium in Fermi levels. As a result, Al film is positively charged, and the *p*-Si region is negatively charged with a depletion region formed (see the detailed illustration in Fig. S1). Therefore, due to these two functionalities, this Si-Al nanostructure can function as a nanoscopic color sensitive photodiode with size-tunable spectral selectivity.

Due to the rotational symmetry of nanodisks, the photo-response is independent of the incident light polarization. Nevertheless, introducing polarization dependence to the photodetectors by symmetry breaking would be straightforward.[34-36] FDTD simulations were performed to investigate the absorption characteristic inside Si region as a function of nanodisk diameter, as shown in Fig. 1b. Strong size-dependent absorption peaks could be engineered to achieve color sensitivity, in stark contrast to the intrinsic material absorption of Si, which monotonically decreases from 400 to 800 nm.[37, 38] In addition, Fig. 1b shows that the fundamental absorption peak is observed at ~400 nm for the nanodisk diameter *D*=50 nm and it redshifts to ~625 nm when *D* increases to 170 nm. Moreover, the absorption peak due to higher-order mode is also observed for larger nanodisks, appearing at 400 nm for nanodisk diameter *D*=130 nm and red shifting to ~440 nm for *D*=190 nm. For comparison, Fig. 1c and Fig. S2 present the absorption characteristics of Si nanodisk with and without Al film respectively, where a sharper response of absorption spectrum is obtained with Al film. The sharper absorption is due to the enhancement of magnetic dipole (*MD*) in Si by the electric dipole (*ED*) of the Al nanodisk on top.

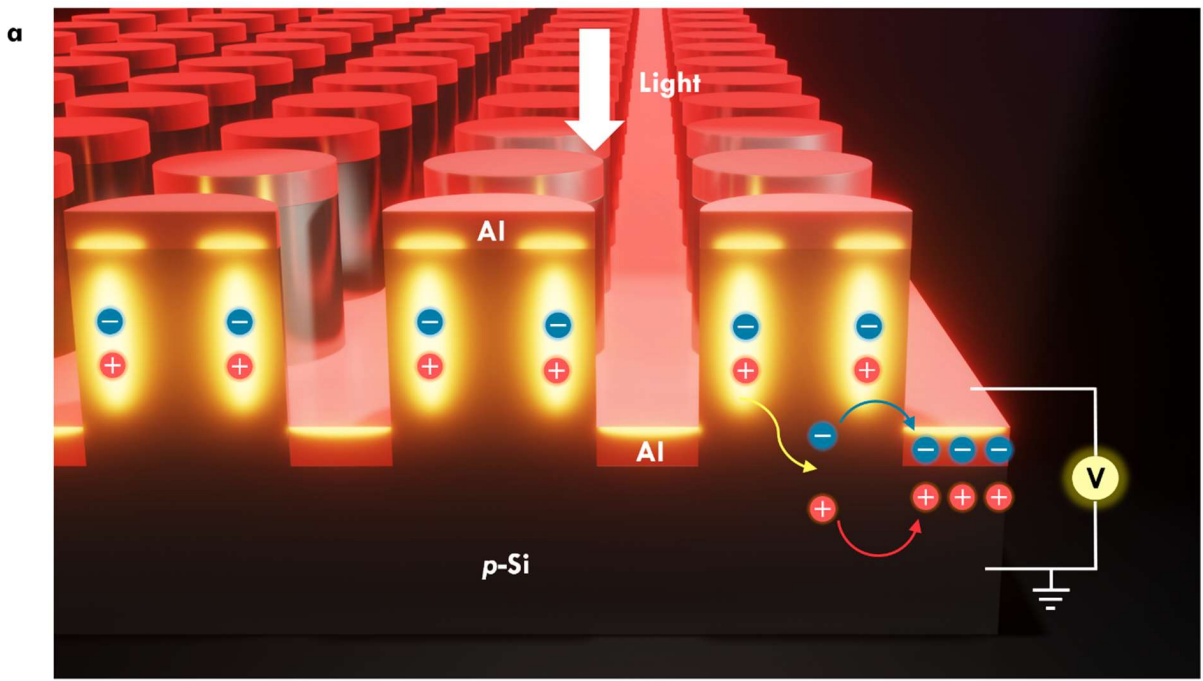

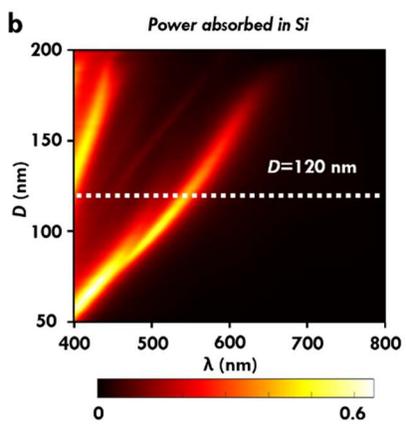
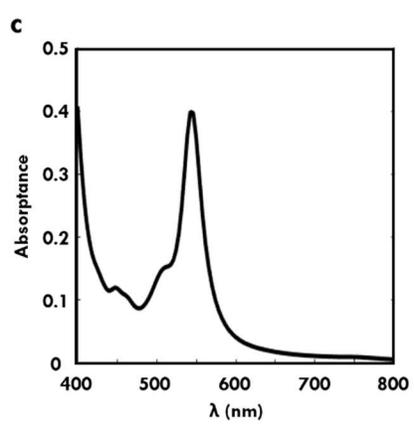
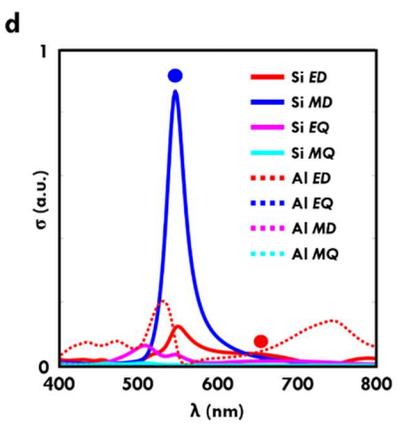

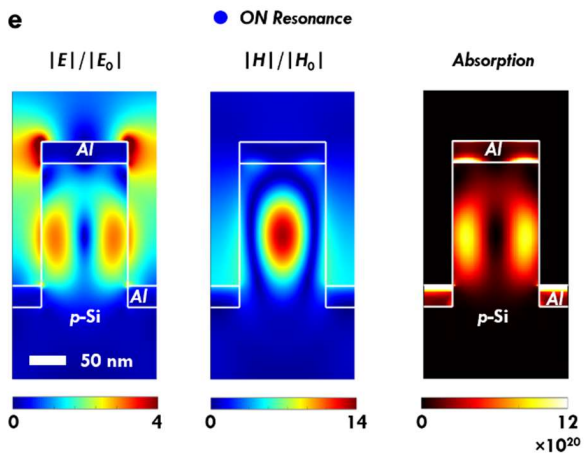
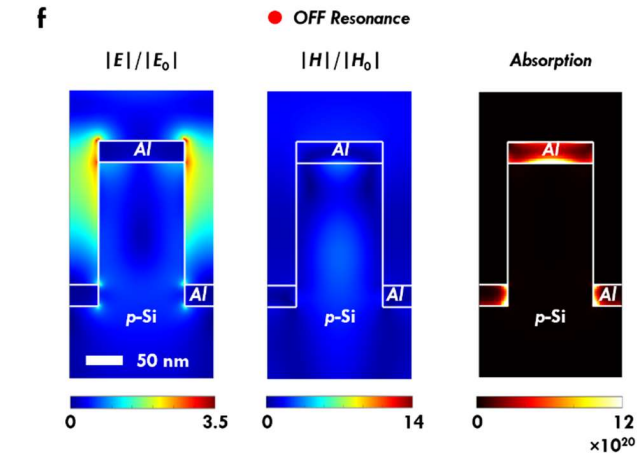

**Figure 1. Design of the color-sensitive optical detectors based on the hybrid *p*-Si and Al nanostructures.** (a) Schematic illustration of the designed color-sensitive detector. The designed device array has a fixed pitch size $\Lambda$ of 200 nm and a height *h* of 200 nm, with 30-nm-thick Al film. (b) Simulated optical absorptance spectra when the Si nanodisk diameter *D* is varied from 50 nm to 200 nm. Higher-order absorption peaks appear for nanodisks with diameters larger than 125 nm. (c) Absorptance spectrum for nanodisk array with a diameter *D*=120 nm. (d) Multipolar decomposition of the scattering cross-section (σ) showing contributions of the electric dipole (*ED*), magnetic dipole (*MD*), electric quadrupole (*EQ*) and magnetic quadrupole (*MQ*) supported by the hybrid Si-Al nanodisk array (*D*=120 nm), highlighting dominant resonance due to *MD* in Si. (e) Spatial distributions of electric field |***E***| and magnetic field |***H***| at the resonance wavelength of ~540 nm, corresponding to the absorption of Si *MD* peak for *D*=120 nm. The absorption plot shows that absorption happens mostly within the *p*-Si nanostructures to generate electron-hole pairs. (f) Spatial distributions of electric field |***E***| and magnetic field |***H***| at an off-resonance wavelength of 650 nm for *D*=120 nm.

To elucidate the nature of the absorption peaks, we show multipolar decomposition results in Fig. 1d. It presents the relative scattering cross-sections of the electric dipole (*ED*), electric quadrupole (*EQ*), magnetic dipole (*MD*) and magnetic quadrupoles (*MQ*) of the Si-Al nanostructure, for *D*=120 nm. Clearly, enhanced absorption is attributed to the magnetic dipole resonance in the Si nanodisk, with corresponding electric field |***E***| and magnetic field |***H***| distributions shown in Fig. 1e. The enhanced electric field around the Al nanodisk is due to the excitation of localized plasmon resonance.[36] The electric field is also highly confined within the Si nanodisk, where it leads to photon absorption and electron-hole pair generation. These generated electron-hole pairs will be charge separated first due to the internal built-in electrical field, and then they diffuse a short distance to be collected by the respective electrodes, such as Al metal at the base of the Si pillars, where the Al film is electrically connected to external electronics. As shown in Fig. 1e, the distance between the peak absorption location inside *p*-Si nanodisks and the Al film is only ~70 nm, where this distance is much smaller than the typical diffusion length of ~100 μm in Si.

On the other hand, the Al film on top of the Si nanodisk is electrically floating, however, a charge depletion region still forms (see the detailed schematic illustration on this junction creation process in Fig. S1). In comparison, Fig. 1f presents the corresponding electric field $|E|$ and magnetic field $|H|$ distribution off-resonance (*i.e.*, $\lambda$=650 nm). It shows that the electrical fields predominantly lie outside the Si nanostructures, leading to minimal light absorption or photocurrent generation. Similar analysis for a larger nanodisk with $d$ = 150 nm shows a more complex field distribution within the nanostructure as the resonant wavelength redshifts to ~600 nm, as shown in Fig. S3.

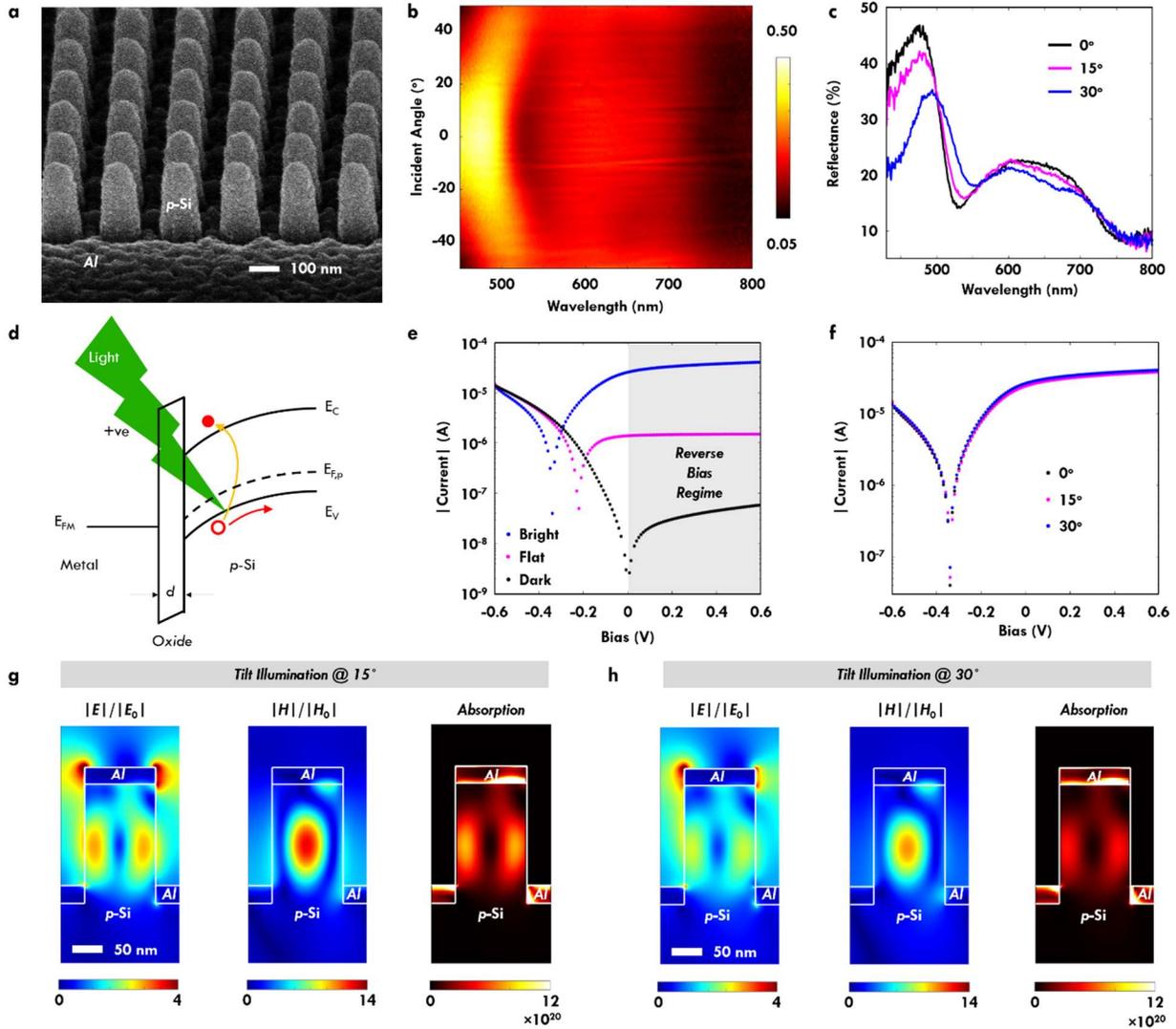

**Figure 2. Characterization results of the fabricated color-sensitive detector devices.** (a) Scanning electron micrograph (SEM) image of a sample patterned with nanodisks with diameter *D* of ~120 nm, a pitch of 200 nm and a height of 200 nm. (b) Angle-resolved reflectance spectra as measured in the back focal plane showing minimal incident angle dependence. (c) Measured reflectance spectra at the respective incident angles of 0º, 15º and 30º. (d) Energy band diagram of the device under reverse bias condition, which widens the depletion region and facilitates charge separation. Electron-hole pairs are generated in the depletion region, followed by charge separation with holes (electrons) moving to the right (left) into *p*-Si semiconductor (metal contact), resulting

in a larger positive current under reverse bias. The thickness $d$ of the native silicon oxide is ~2 nm based on ellipsometer measurement. (e) *I-V* characterization result of the nanostructured device at logarithmic scale under "dark" and "bright" illumination conditions. The observed *I-V* curves correspond to a typical Schottky barrier photodiode device, where photocurrent is clearly observed under reverse bias condition (*i.e.*, a positive voltage being applied to the top Al electrode and Si back contact set to ground). The incident light source is a 532 nm laser with a laser power of 0.78 mW, where the photocurrent measured from the flat un-patterned region is shown for comparison. (f) Photocurrent measurement of the detector element with the respective tilt angles of 0º, 15º and 30º. It shows that the photocurrent from the nanostructured color detector does not change with respect to the tilted light illuminations. (g)-(h) Simulated absorption profiles when the laser illumination has the respective tilt angles of 15º and 30º.

To fabricate the samples, we exposed nanodisk patterns using electron beam lithography in hydrogen silsesquioxane (HSQ) resist that was spin-coated on *p*-type prime grade <100> Si wafers (Si Valley Microelectronics Inc., resistivity of 10-50 ohm-cm). Samples were then etched using inductively coupled plasma (ICP) to form Si nanodisk arrays[39] (see details in the Methods section). The thickness of HSQ resist is optimized such that HSQ mask is etched away completely when the dry etching is done. After the etching process, 30-nm-thick Al film was then evaporated onto the sample to work as the top electrode via electron-beam evaporation, to form a reliable Si-Al Schottky junction. We observed that HF treatment should not be performed on these Si nanostructures, as it probably introduces additional surface states that produce Ohmic-like contact between Si and Al (see Fig. S4 for details).

Figure 2a presents the scanning electron microscope (SEM) image of the Si-Al nanoantenna with a diameter $D$ of 120 nm. The corresponding angle-resolved reflectance spectrum in the back focal plane are shown in Fig. 2b and Fig. 2c presents the reflectance spectra at the respective incident angles of 0°, 15°, and 30°. To characterize the Schottky barrier photodiode, the diode is reverse biased as shown in Fig. 2d, where a positive voltage is applied to the top Al electrode with the Si back contact set to ground. Fig. 2e presents the measured $I$-$V$ curves, which were obtained from a Keithley 2450 sourcemeter with and without laser illumination. The corresponding $I$-$V$ curves at linear scale are shown in Fig. S5, showing a dark current of ~52 nA at a reverse bias voltage of 0.5 V.

The electrical characteristics of the nanostructured Al-Si Schottky device can be modelled using equations for thermionic emission:[40, 41]

$$I = AA^{**}T^2 \exp\left(-\frac{q\phi}{kT}\right)\left[\exp\left(\frac{qV}{nkT}\right) - 1\right], \qquad (1)$$

where $q$ denotes the fundamental charge of an electron, $T$ is the device temperature, $k$ is the Boltzmann constant, $A$ is the device electrode area of ~0.07 cm$^2$, and $A^{**}$ denotes the Richardson constant of $p$-Si =32 Acm$^{-2}$K$^{-2}$. When fitting the $I$-$V$ curve in the linear range of 0.04 V to 0.10 V based on Eq. (1), the Schottky barrier height (SBH) $\phi$ of the device is estimated to be ~0.816 eV with an ideality factor $n$ of ~1.09.

In addition, the depletion width $W$ of the Schottky barrier is estimated using:

$$W = \sqrt{\frac{2\epsilon_s}{qN_A}\left(V_{Bi} - V - \frac{kT}{q}\right)}, \qquad (2)$$

where $V_{Bi}$ denotes the built-in-potential being the difference between the Schottky barrier height and the Fermi level of the semiconductor. $V_{Bi}$ has a value of 0.587 V, based on the exacted Schottky barrier height $\phi$ of ~0.816 eV.[42] $\epsilon_s$ denotes the relative permittivity and $\epsilon_s$=11.68.[41] $N_A$ denotes the doping concentration of the $p$-Si, with a value of ~1.38×10$^{15}$ cm$^{-3}$. $V$ denotes the externally applied potential. The depletion width $W$ is calculated to be ~700 nm when the externally applied potential $V$ is 0, and it means that the $p$-Si nanostructures are completely depleted.

The measured responsivity of the nanostructured detector element is 47.2 mA/W, corresponding to an external quantum efficiency of ~9%. At this bias condition and with 1 mW of 532 nm laser illumination, the photocurrent was ~800-fold higher than the dark current. Moreover, the photocurrent was ~31-fold higher with laser illuminating the patterned array compared to the flat region, as shown in Fig. 2e. In addition, Fig. 2f presents photocurrent measurement of the detector element with the respective tilt angles of 0º, 15º and 30º. It shows that the photocurrent from the nanostructured color detector does not change with respect to the incident angle, corroborated with simulated absorption profiles for incident angles of 15º and 30º in Fig. 2g-h that look similar to that of normal incidence. In other words, the optical antenna effects of the hybrid Si-Al nanostructures are having similar functionalities of micro-lenses, which are able to collect the light even for the inclined illuminations.

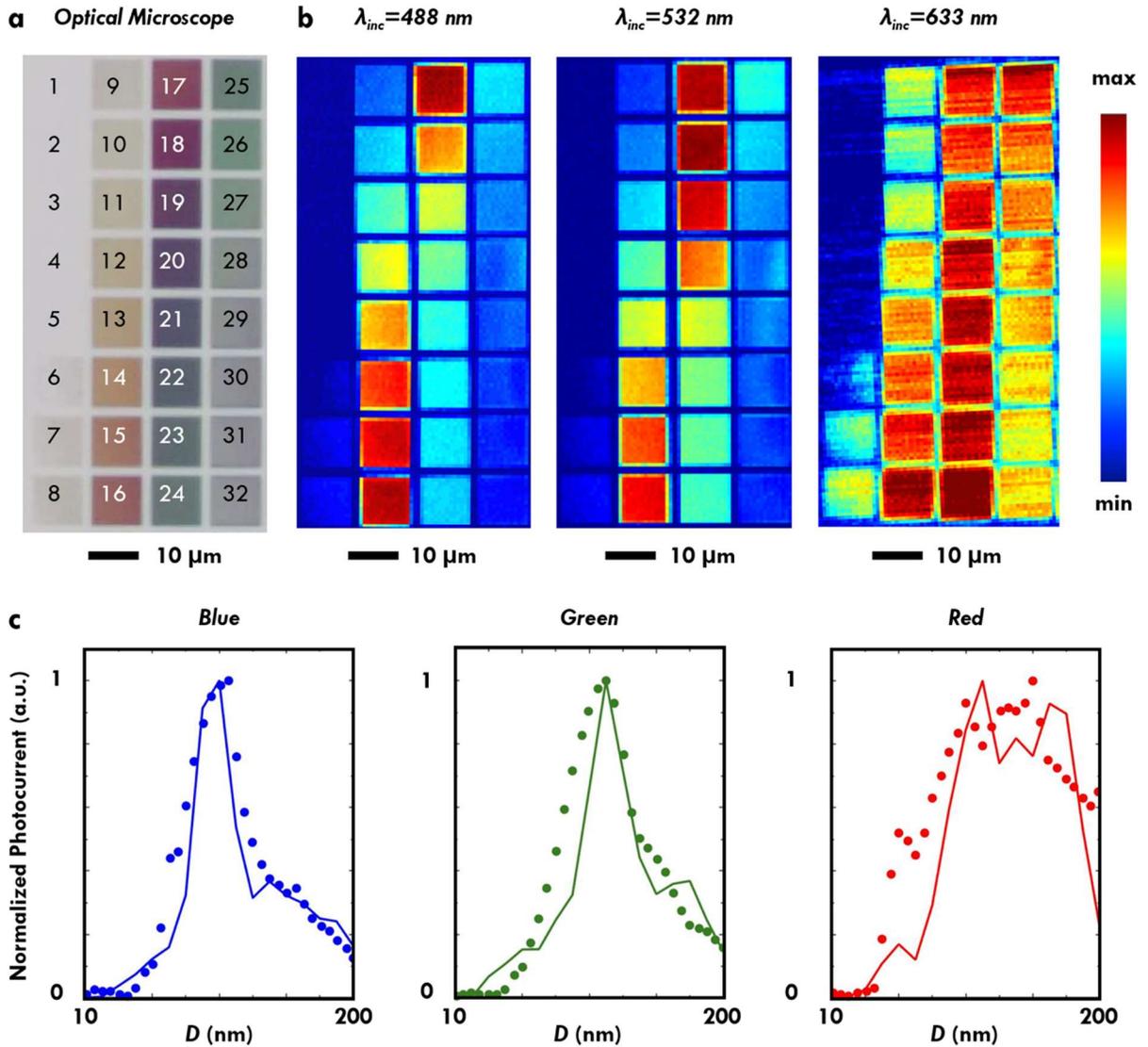

**Figure 3. Optical and photocurrent measurements at red, green and blue wavelengths for arrays of Si-Al hybrid structures.** (a) Optical microscope reflection image of the device. It consists of 32 color sensitive photodetector arrays with diameter $D$ ranging from 10 to 200 nm. Each photodetector array has an area of $10\times10$ μm$^2$. (b) Normalized photocurrent maps of the device under the illumination laser with the respective wavelengths of 488 nm (left), 532 nm (center) and 633 nm (right). The device was reversed biased at 0.5 V. (c) Comparison of the

experimentally observed photocurrent (dots) with simulated absorption in the Si region (solid curves), showing good agreement. The simulation was carried out with 50×50 nanodisks.

To characterize the wavelength-dependent photo-response of the nanodisks systematically, we tested 32 nanodisk arrays with a constant period of 200 nm and the diameter $D$ varied from 10 to 200 nm in steps of ~5.4 nm, as indicated in the optical micrograph in Fig. 3a. Meanwhile, the nanodisk arrays with diameter $D$ of less than 50 nm are not fully fabricated since some of these nanodisks with small diameters do not have sufficient adhesion to the substrate during the resist development process. For optical characterization, we raster-scanned CW lasers with wavelengths of 488 nm (blue), 532 nm (green), and 633 nm (red) over the device and recorded the photocurrent at each position of the sample. The device was reverse biased at 0.5 V, and the intensity of the lasers was kept at 0.3 mW. Figure 3b shows the normalized photocurrent maps of the arrays, and wavelength-dependent photo-response is clearly observed as the arrays with larger nanodisk diameters have higher photocurrent as the wavelength is increased. A generally good agreement was observed between the experimentally measured photocurrent and the simulated absorption of the Si region, as shown in Fig. 3c.

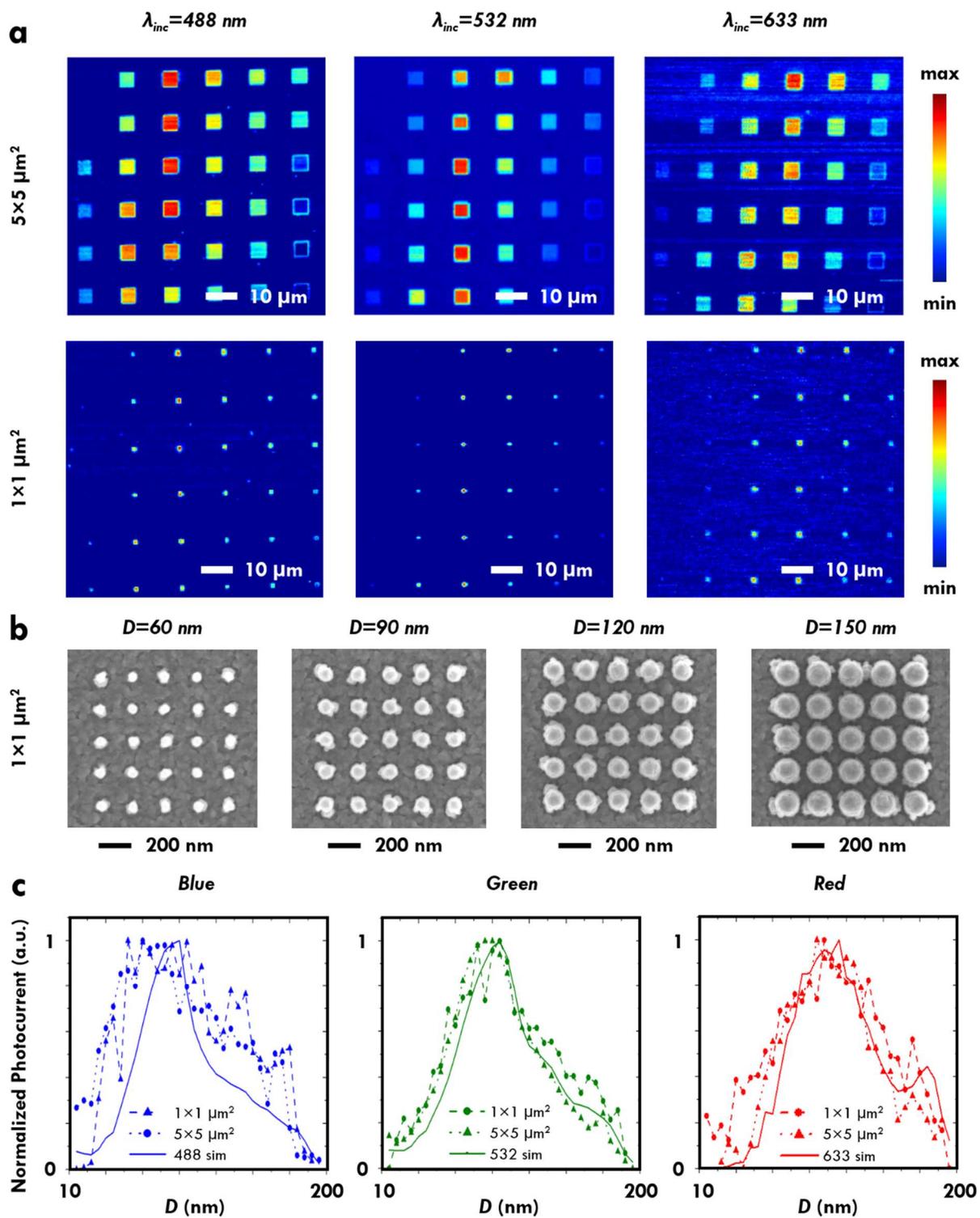

**Figure 4. Miniaturizing the color-sensitive detector devices down to 1×1 μm².** (a) Normalized photocurrent maps of the device under 488 nm (left), 532 nm (center) and 633 nm (right) laser

illumination for the detector devices with the respective dimensions of 5×5 μm² and 1×1 μm². The device consists of 36 color-sensitive photodetector arrays and was reverse biased at 0.5 V. The nanodisk diameters were varied nominally from 10 nm to 200 nm in these devices, where the nanodisk height is 200 nm and the pitch size is fixed at 200 nm. (b) SEM images of the typical nanodisk array with the area of 1×1 μm². (c) Experimentally measured photocurrent response (dots) from the devices with the area of 1×1 μm² and 5×5 μm², in comparison with simulation result (solid curves), showing a good agreement. The simulation was carried out with 5×5 nanodisks.

Next, we reduced the dimensions of these detector devices from 10×10 μm², down to 5×5 μm² and 1×1 μm². Figure 4a shows the spectral response of three representative arrays sensitive to red, green, and blue. These device arrays consist of 36 color-sensitive photodetector devices, where the nanodisk arrays have a height of 200 nm, a fixed period of 200 nm and a varying diameter $D$ from 10 nm to 200 nm. Generally, the peak resonance wavelength is red-shifting when the disk diameter increases. Figure 4b presents the SEM images of the typical nanodisk arrays with an area of 1×1 μm², while the SEM images of the typical nanodisk arrays with an area of 5×5 μm² are shown in Fig. S6. The corresponding photocurrent response (dots) from the devices as measured experimentally are shown in Fig. 4c. These measurement results are benchmarked with simulation results (solid curves), showing a good agreement.

**Discussions**

The miniaturized color-sensitive detectors could be potentially made into a single-element design as shown in Fig. S7, where each detector has the accompanying electronics with a couple of transistors connected to it, similar to the current CMOS sensors. In addition, this color-sensitive photodetector can be extended to achieve polarization sensitivity via asymmetric nanostructures, such as ellipses or dimers.[35] Moreover, the spectral range of our current detector design could be potentially extended into the ultraviolet (UV) regime by exploring Si interband plasmonics.[33]

**Conclusion**

We have designed and fabricated color-sensitive hybrid Si-Al photodetectors with a small pixel dimension of 1×1 μm$^2$. This hybrid Si-Al nanostructure is performing two functionalities at the same time. First, it has the hybrid Mie-plasmon resonance, especially the magnetic dipole component, to achieve the spectrally selective absorption of light, which is consequently converted into electron-hole pairs. At the same time, this hybrid Si-Al nanostructure is essentially forming a Schottky barrier, which can separate the generated electron-hole pairs for photodetection. The technology developed herein could be potentially used to replace the traditional dye-based CMOS camera filters for easier fabrication of camera sensors at higher pixel densities as well as pushing detection range to UV regime.[33]

**METHODS**

**Fabrication of Si Nanoantenna Array.** Hydrogen silsesquioxane (HSQ) etching mask was first fabricated on the single crystalline *p*-doped Si (resistivity = 10-50 Ω/cm, prime grade, Si Valley Microelectronics, Inc.). HSQ resist (Dow Corning XR-1541-006) was first diluted to 3% by using methyl isobutyl ketone (MIBK) solvent and spin-coated onto a cleaned substrate at 3k round-per-minute (rpm) to obtain a HSQ thickness of ~50 nm.[43] Electron beam exposure was then carried out with an electron acceleration voltage of 100 keV (EBL, Elionix ELS-7000), beam current of 500 pA, and an exposure dose of ~12 mC/cm$^2$. The sample was then developed by NaOH/NaCl salty solution (1% wt./4% wt. in de-ionized water) for 60 seconds and then immersed in de-ionized water for 60 seconds to stop the development. The sample was immediately rinsed by acetone, isopropanol alcohol (IPA) and dried by a continuous nitrogen gas flow. Si etching of 200 nm was then carried out by inductively-coupled-plasma (ICP, Oxford Instruments Plasmalab System 100),[43] with an RF power of 100 watts, ICP power of 150 watts, Cl$_2$ with a flow rate of 22 sccm (standard-cubic-centimeters-per-minute), under a process pressure of 5 mTorr, and temperature of 40 °C.

**Scanning Electron Microscope (SEM).** SEM images were taken at an acceleration voltage of 1 keV (Hitachi, SU8220).

**Optical and Electrical Characterizations.** Keithley 2450 source meter was used to obtain the *I-V* characteristics. The electrode on the surface of the sample that has the nanoantenna arrays is connected to the high potential terminal, 'Force-Hi', of the source meter, while the back surface of the sample is connected to the low potential terminal, 'Force-Lo'. Before connecting the sample to the source meter, the back surface of the sample is lightly scraped to remove the native oxide layer. For the photocurrent measurements, our light source is a super-continuum laser (*i.e.*, NKT Photonics, SuperK Compact), which emits a broad spectrum ranging from 400 nm to 2400 nm. The output of the super-continuum laser is then passed through a monochromator (*i.e.*, NKT Photonics. SuperK VARIA), which can select the wavelength from 430 to 680 nm with a step size of 10 nm. The monochromatic light is then focused onto the nanoantenna array device with a 50× objective lens that has a numerical aperture (N.A.) of 0.50. For each wavelength, the photocurrent generated by the nanoantenna array is measured with the Keithley 2450 source meter. During the photocurrent measurements, the nanoantenna array is reverse-biased at 0.50 V. The detailed formula for calculating the external quantum efficiency and responsivity is shown in the supporting information Section S8.

**Numerical Simulations.** Finite-difference time-domain (FDTD) simulations of the reflectance spectra and electric field distribution were carried out by using Lumerical FDTD Solutions. Periodic boundary conditions were used with the incident optical field being linearly polarized and at different incident angles. In addition, the details of multipolar decomposition could be found in the reference.[44, 45]

**AUTHOR INFORMATION**

**Corresponding Authors**

*E-mail: joel_yang@sutd.edu.sg. Telephone: +65 64994767

*Email: dongz@imre.a-star.edu.sg. Telephone: +65 63194857

**ORCID**

Jinfa Ho: 0000-0001-6884-4785

Zhaogang Dong: 0000-0002-0929-7723

Soroosh Daqiqeh Rezaei: 0000-0002-9807-2074

Arseniy I. Kuznetsov: 0000-0002-7622-8939

Joel K. W. Yang: 0000-0003-3301-1040

**Author contributions**

J.H., Z.D. and J.K.W.Y. conceived the concept, designed the experiments and wrote the manuscript. H.S.L., J.H., J.Z. and Z.D. performed the *I-V* electrical measurements and optical characterizations. F.T., J.H. and Z.D. performed the nanofabrication processes. K.C.H.G. and M.W. performed the evaporation of Al films. J.H. and S.D.R. did the numerical simulations. S.L. and F.T. performed the initial fabrication of electrodes. J.C. helped to setup the *I-V* characterization platform. C.P.Y.W. helped on the HF treatment and the model fitting of Schottky barrier. A.I.K. participated in discussions and gave suggestions. All authors analyzed the data, read and corrected the manuscript before the submission. J.H. and Z.D contributed equally to this work.

**Notes**

The authors declare no competing financial interests.

**ACKNOWLEDGMENTS**

We would like to acknowledge the funding support from Agency for Science, Technology and Research (A*STAR) SERC Pharos project (grant number 1527300025) and MTC Programmatic grant number M21J9b0085. In addition, Z.D. and J.K.W.Y. acknowledges the funding support of A*STAR AME IRG (Project No. A20E5c0093). Z.D. would like to acknowledge the funding support from A*STAR Career Development Award (CDA) grant (Project No. C210112019), A*STAR MTC IRG (Project No. M21K2c0116) and A*STAR MTC YIRG (Project No. M21K3c0127). C.P.Y.W. acknowledges funding support from A*STAR SERC Pharos Project (grant number 1527000016) and A*STAR AME YIRG funding (Project No. A2084c0179).

*Supplementary Information*

# Miniaturizing Color-Sensitive Photodetectors via Hybrid Nanoantennas towards Sub-micron Dimensions

Jinfa Ho[1,†], Zhaogang Dong[1,2,†,*], Hai Sheng Leong[1], Jun Zhang[1], Febiana Tjiptoharsono[1], Ken Choon Hwa Goh[1], Soroosh Daqiqeh Rezaei[3], Mengfei Wu[1], Shiqiang Li[1], Jingyee Chee[1], Calvin Pei Yu Wong[1], Arseniy I. Kuznetsov[1], and Joel K. W. Yang[1,3,*]

[1]Institute of Materials Research and Engineering, A*STAR (Agency for Science, Technology and Research), 2 Fusionopolis Way, #08-03 Innovis, 138634 Singapore

[2]Department of Materials Science and Engineering, National University of Singapore, 9 Engineering Drive 1, 117575, Singapore

[3]Singapore University of Technology and Design, 8 Somapah Road, 487372, Singapore

*Correspondence and requests for materials should be addressed to J.K.W.Y. (email: joel_yang@sutd.edu.sg) and Z.D. (email: dongz@imre.a-star.edu.sg).

## S1. Charge distribution of Schottky Barrier as formed by *p*-Si and Al.

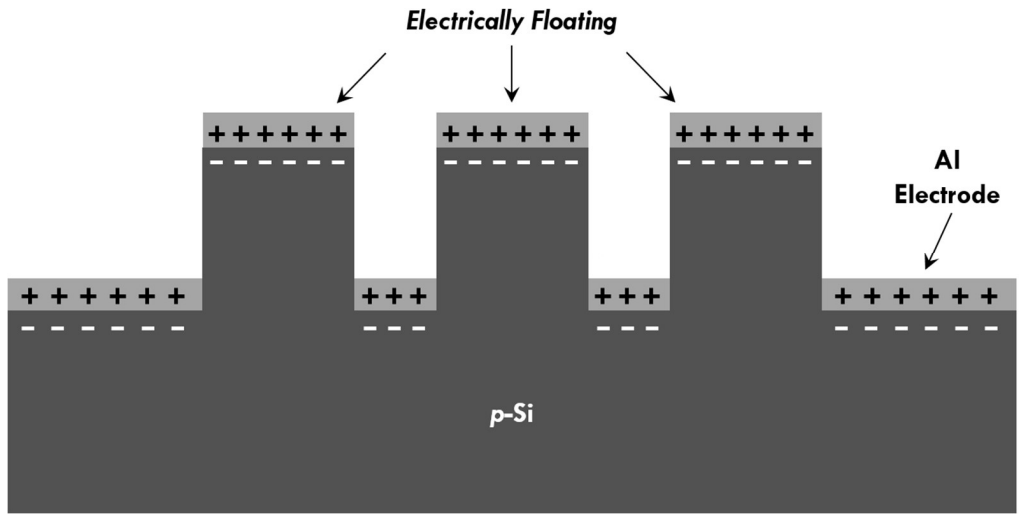

**Figure S1. Schematic illustration on the static charge distribution on the Schottky barrier as formed by *p*-Si and Al.** The holes in *p*-Si will diffuse onto the surface of Al film to equilibrate fermi levels, and a depletion region will be formed in the *p*-Si region, where the *p*-Si region will be negatively charged and Al film will be positively charged. In addition, this schematic shows that the Al film on top of the *p*-Si nanodisks is electrically floating. Instead, the Al film at the flat region is electrically connected to the source meter.

## S2. Simulation analysis on the silicon nanodisks without aluminum film.

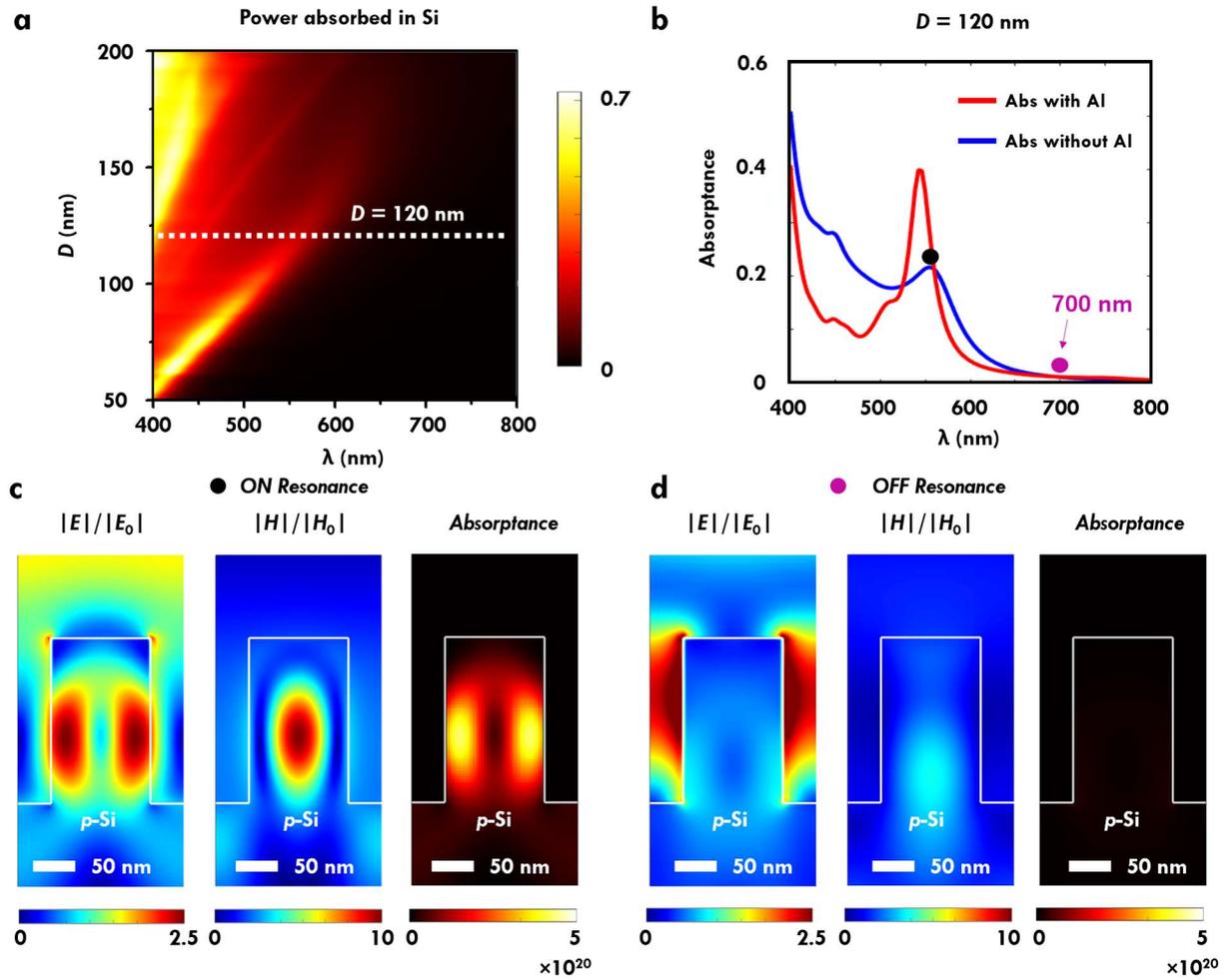

**Figure S2. Simulation analysis on the silicon nanodisks without aluminum film.** (a) Simulated optical absorptance spectra when the Si nanodisk diameter *D* is varied from 50 nm to 200 nm, with a fixed pitch size *Λ* of 200 nm and a height *h* of 200 nm. (b) Absorptance spectra for silicon nanodisk (*D*=100 nm) with and without 30-nm-thick aluminum film. It shows that the hybrid Si-Al nanostructure has a sharper absorptance spectrum. (c)-(d) Spatial distributions of electric field |***E***|, magnetic field |***H***| and absorptance at the respective conditions of ON resonance (*i.e.* 550 nm) and OFF resonance (*i.e.*, 700 nm).

## S3. Simulation analysis on the nanodisk with the diameter of 150 nm.

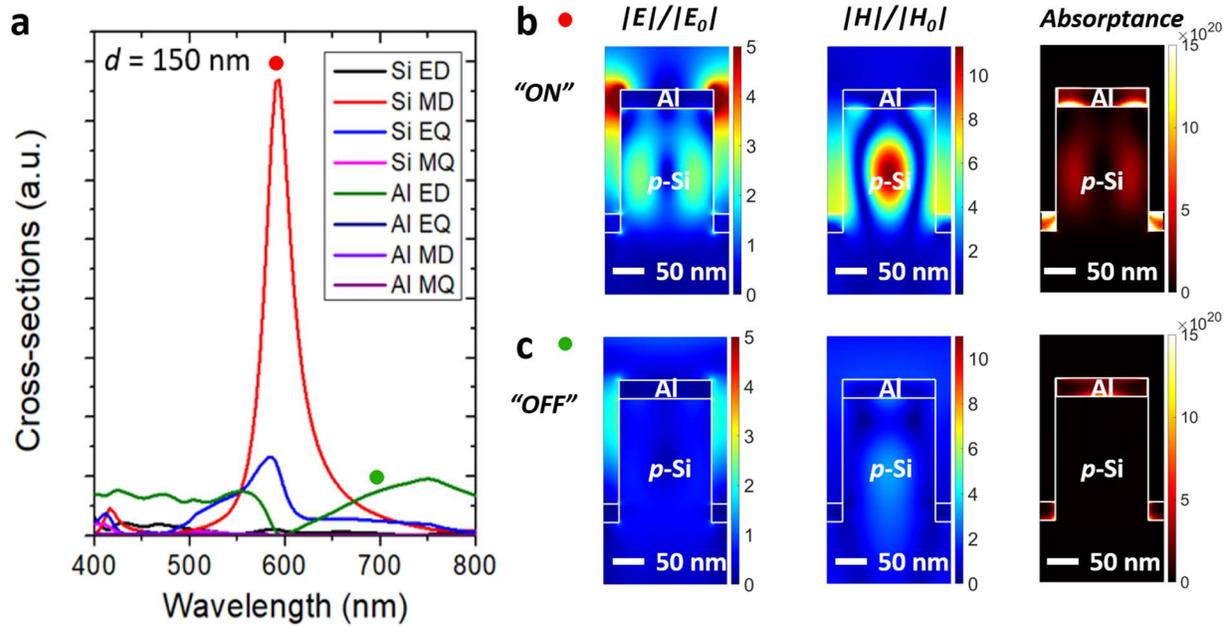

**Figure S3. Simulation analysis on the nanodisk with the diameter of 150 nm.** (a) Scattering cross-section of the electric dipoles (*ED*), magnetic dipoles (*MD*), electric quadrupoles (*EQ*) and magnetic quadrupoles (*MQ*) supported by the silicon and aluminum nanodisks with a diameter $D$ of 150 nm. The designed device array has a fixed pitch size $\Lambda$ of 200 nm and a height $h$ of 200 nm, with the evaporated aluminum film thickness of 30 nm. (b)-(c) Spatial distributions of electric field $|E|/|E_0|$ and magnetic field $|H|/|H_0|$ at the respective wavelengths of 600 nm (b) and 700 nm (c).

**S4. *I-V* characterization result of the Si-Al device after the HF treatment.**

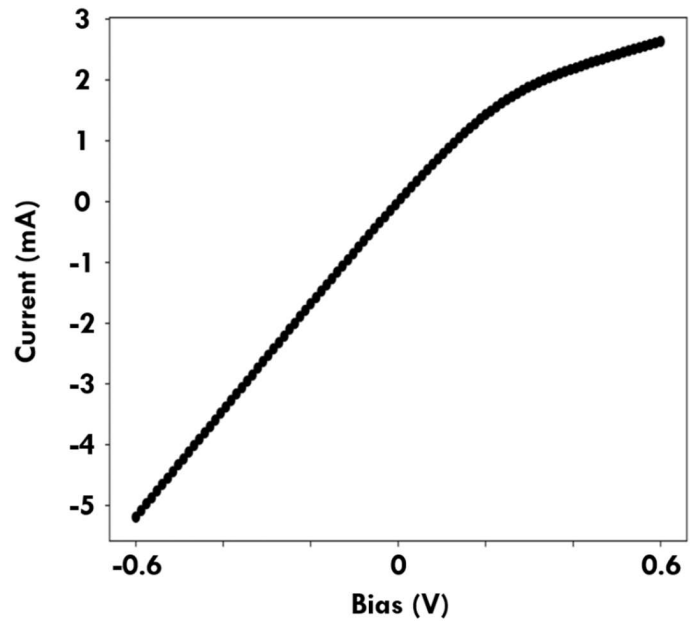

**Figure S4. *I-V* characterization result of the Si-Al device after the surface treatment of 1% hydrofluoric (HF) acid in DI water under bright illumination conditions.** This plot is based on the linear scale, and it shows that this Si-Al junction becomes an Ohmic contact after the HF treatment.

## S5. Linear plot of the *I-V* characterization results for the Si-Al device.

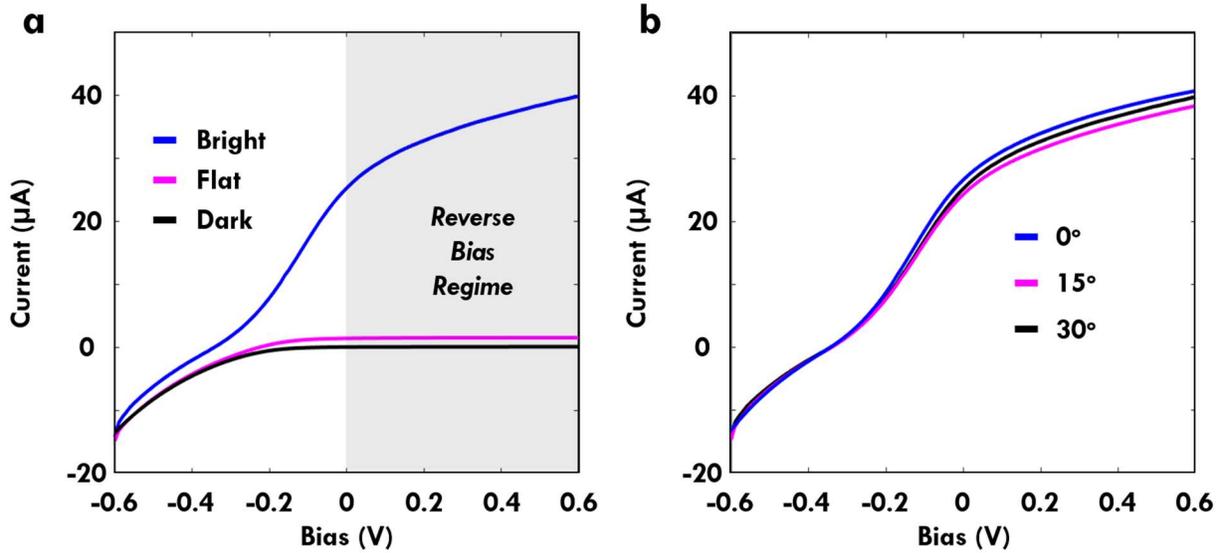

**Figure S5. Linear plot of the *I-V* characterization results for the Si-Al device.** (a) *I-V* characterization result of the nanostructured device at the linear scale under dark and bright illumination conditions. "Flat" denotes the current as measured when light illuminates the flat region of the sample. (b) Photocurrent measurement of the detector element with the respective tilt angles of 0°, 15° and 30°. The corresponding plots in log scale are shown in Fig. 2.

**S6. SEM images of the typical nanodisk arrays with the area of 5×5 μm².**

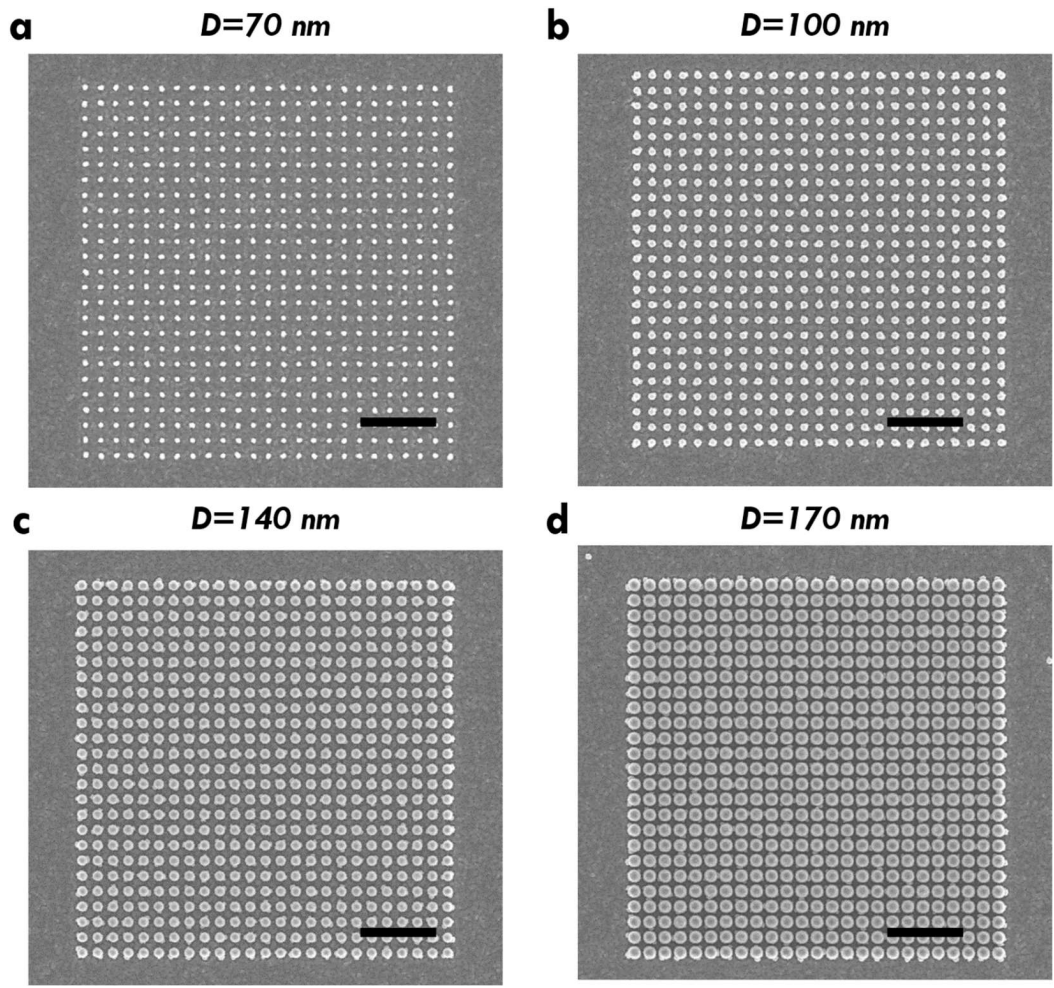

**Figure S6. SEM images of the typical nanodisk arrays with the area of 5×5 μm².** (a) *D*=70 nm. (b) *D*=100 nm. (c) *D*=140 nm. (d) *D*=170 nm. The scale bars denote 1 μm.

**S7. Potential single-element optical detector with the electrical readout circuit.**

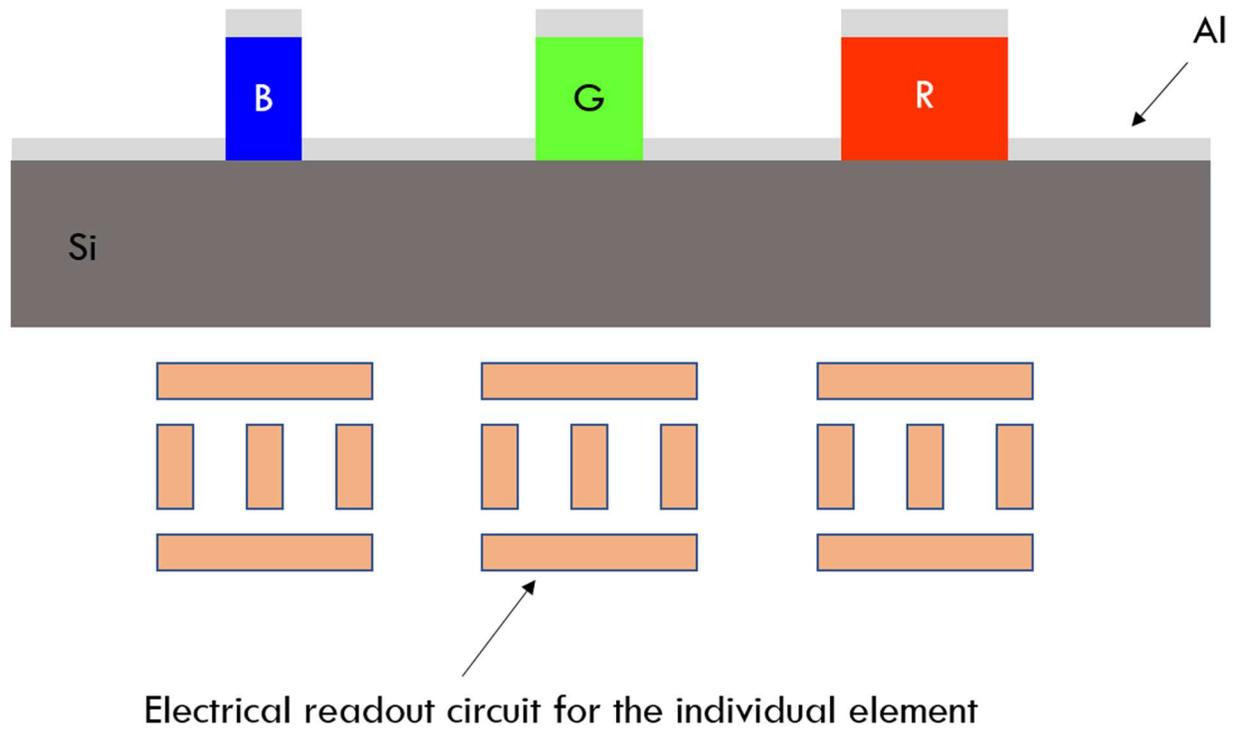

**Figure S7. Potential single-element optical detector with the electrical readout circuit.** The individual R-G-B detector elements with different dimensions are labelled as "R", "G" and "B". The electrical readout circuit based on electrical interconnects could be fabricated based on the CMOS technologies.[1, 2]

## S8. Formula for calculating external quantum efficiency and responsivity.

The external quantum efficiency $\eta$ is defined as the ratio between the number of output electrons and the number of incident photons:[3]

$$\eta = \frac{i_d/e}{P/h\nu}, \tag{S1}$$

where $i_d$ denotes the photocurrent of the detector. $e$ denotes the fundamental charge of an electron. $P$ denotes the incident laser power. $h$ denotes the Planck constant. $\nu$ denotes the light frequency.

In addition, the responsivity $\hat{R}$ is defined as the ratio between the output current and the input optical power:

$$\hat{R} = \frac{i_d}{P}. \tag{S2}$$

Therefore, the relationship between the external quantum efficiency $\eta$ and the responsivity $\hat{R}$ is given by:

$$\eta = \hat{R}\frac{h\nu}{e}. \tag{S3}$$

If we put all the numbers inside, Eq. S3 could be simplified into:

$$\hat{R} = \eta \frac{\lambda(\mu m)}{1.24}, \tag{S4}$$

where $\lambda$ has the unit of μm and $\hat{R}$ has the unit of A/W.